\magnification = 1200
\overfullrule=0pt
\def\l{{\langle}}
\def\r{{\rangle}}
\def\cov{{\rm cov}}
\def\var{{\rm var}}
\def\p{{\partial}}
\def\C{{\cal C}}
\def\B{{\cal B}}

\centerline{\bf $\chi^2$ and Linear Fits}
\bigskip
\centerline{Andrew Gould (Dept.\ of Astronomy, Ohio State University)}
\bigskip
\bigskip
\centerline{\bf Abstract}
\bigskip
The mathematics of linear fits is presented in covariant form.  Topics
include: correlated data, covariance matrices, joint fits to multiple
data sets, constraints, and extension of the formalism to non-linear fits.
A brief summary at the end provides a convenient crib sheet.
These are somewhat amplified notes from a 90 minute lecture in a first-year
graduate course.  None of the results are new.  They are presented here
because they do not appear to be elsewhere available in compact form.
\bigskip
\bigskip
\centerline{\bf Expectations and Covariances}
\bigskip
Let $y$ be random variable, which is drawn from a random distribution
$g(y)$.  Then, we define the ``expected value'' or ``mean'' of $y$ as
$$
\l y\r \equiv \int y g(y) dy \bigg/\int g(y) dy.
$$
Using this definition, it is straightforward to prove the following
identities,
$$
\l y_1 + y_2 \r = \l y_1 \r + \l y_2 \r,\quad \l k y \r = k\l y \r,
\quad \l k \r = k,\quad \l\l y \r\r = \l y \r,
$$
where $k$ is a constant.  We now motivate the idea of a ``covariance''
of two random variables by noting
$$
\l y_1 y_2 \r = \l y_1 \r \l y_2 \r + \cov(y_1,y_2)
$$
where
$$\cov(y_1,y_2) \equiv \l (y_1 - \l y_1 \r)(y_2 - \l y_2 \r)\r
= \l y_1y_2 \r - \l y_1 \r \l y_2 \r,
$$
the last step following from the identities above.  If, when $y_1$ is
above its mean, then $y_2$ also tends to be above its mean (and similarly for
below), then $\cov(y_1,y_2)>0$, and then $y_1$ is said to be ``correlated''
with $y_2$.  If, when $y_1$ is above then $y_2$ is below, then  
$\cov(y_1,y_2)<0$, and the $y_1$ and $y_2$ are said to be ``anti-correlated''.
If $\cov(y_1,y_2)=0$, they are said to be ``uncorrelated''.  Only in this
case is it true that $\l y_1y_2 \r = \l y_1 \r \l y_2 \r $.
Three other identities that are easily proven,
$$
\cov(ky_1,y_2) = k\cdot\cov(y_1,y_2), \quad 
\cov(y_1,y_2+y_3) = \cov(y_1,y_2)+ \cov(y_1,y_3), \quad
\cov(y,k)=0.
$$
The covariance of a random variable with itself
is called its variance.  The error, 
$\sigma$, is defined to be the square root of the variance,
$$
\var(y) \equiv \cov(y,y) = \l y^2 \r - \l y \r^2,\qquad 
\sigma(y)\equiv \sqrt{\var(y)}
$$
\bigskip
\centerline{\bf Definition of $\chi^2$}
\bigskip
Suppose I have a set of $N$ data points $y_k$ with associated (uncorrelated 
for now) errors $\sigma_k$, and I have a model that makes predictions
for the values of these data points $y_{k,\rm mod}$.  Then $\chi^2$ is
defined to be
$$
\chi^2 \equiv \sum_{k=1}^N {(y_k - y_{k,\rm mod})^2\over \sigma_k^2}.
$$
If the errors are Gaussian, then the likelihood is given by 
${\cal L} = \exp(-\chi^2/2)$, so that
minimizing $\chi^2$ is equivalent to maximizing ${\cal L}$.  However,
even if the errors are not Gaussian, $\chi^2$ minimization is a well-defined
procedure, and none of the results given below (with one explicitedly noted
exception) depend in any way on the errors being Gaussian.  This is important
because sometimes little is known about the error distributions beyond their
variances.

More generally, the errors might be correlated.  Although in practice
this is the exception, it makes the math much easier to consider the
more general case of correlated errors.  In this case, the covariance
matrix, $\C_{kl}$, of the correlated errors (and its inverse $\B_{kl}$)
are defined by
$$
\C_{kl} \equiv \cov(y_k,y_l),\qquad \B\equiv \C^{-1},
$$
Both $\C$ and $\B$ are symmetric.  Then $\chi^2$ is written as
$$
\chi^2 = \sum_{k=1}^N \sum_{l=1}^N (y_k - y_{k,\rm mod})\B_{kl}
(y_l - y_{l,\rm mod})
$$
Note that for the special case of uncorrelated errors, 
$\C_{kl} = \delta_{kl}\sigma_k^2$, where the Kronecker delta is defined by
$$
\delta_{kl} = 1 \quad (k=l),\qquad \delta_{kl} = 0 \quad (k\not=l).
$$
In this case, $\B_{kl} = \delta_{kl}\sigma_k^{-2}$, so
$$
\chi^2 = \sum_{k=1}^N \sum_{l=1}^N 
(y_k - y_{k,\rm mod}){\delta_{kl}\over \sigma_k^2}(y_l - y_{l,\rm mod})
= \sum_{k=1}^N {
(y_k - y_{k,\rm mod})(y_k - y_{k,\rm mod})\over \sigma_k^2},
$$
which is the original definition.
\bigskip
\centerline{\bf Linear model}
\bigskip
A linear model of $n$ parameters is given by
$$
y_{\rm mod} \equiv \sum_{i=1}^n a_i f_i(x),
$$
where the $f_i(x)$ are $n$ arbitrary functions of the independent variable $x$.
The independent variable is something that is known exactly (or very
precisely), such as time.  If the independent variable is subject to
significant uncertainty, the approach presented here must be substantially
modified.

We can now write $\chi^2$,
$$
\chi^2 = \sum_{k=1}^N \sum_{l=1}^N 
\bigg[y_k - \sum_{i=1}^n a_i f_i(x_k)\bigg]\B_{kl}
\bigg[y_l - \sum_{j=1}^n a_j f_j(x_l)\bigg].
$$
The proliferation of summation signs is getting annoying.  We can
get rid of all of them using the ``Einstein summation convention'',
whereby we just agree to sum over repeated indices.  In the above cases,
these are $k,l,i,j$. We then write,
$$
\chi^2 = [y_k - a_i f_i(x_k)]\B_{kl}[y_l - a_j f_j(x_l)],
$$
which is a lot simpler.  Note that $k$ and $l$ are summed over the $N$ data 
points, while $i$ and $j$ are summed over the $n$ parameters.
\bigskip
\centerline{\bf Minimizing $\chi^2$}
\bigskip
The general problem of minimizing $\chi^2$ with respect to the parameters
$a_i$ can be difficult, but for linear models it is straightforward.
We first rewrite,
$$
\chi^2 = y_k\B_{kl}y_l - 2 a_i d_i + a_i b_{ij} a_j,
$$
where the $d_i$ and $b_{ij}$ are defined by
$$
d_i\equiv y_k \B_{kl} f_i(x_l),\qquad b_{ij}\equiv f_i(x_k)\B_{kl}f_j(x_l).
$$
We find the minimum by setting all the derivatives of $\chi^2$ with respect
to the parameters equal to zero,
$$
0 = {\p\chi^2\over \p a_m} = -2 \delta_{im} d_i + \delta_{im} b_{ij} a_j
 + a_i b_{ij} \delta_{jm} = -2 d_m + b_{mj}a_j + a_i b_{im}=2(b_{mj}a_j-d_m),
$$
where I have used $\p a_i/\p a_m=\delta_{im}$ and where I summed over
one dummy index ($i$ or $j$) in each step.
This equation is easily solved:
$$
d_m = b_{mj}a_j \Rightarrow a_i = c_{ij} d_j,
$$
where
$c_{ij}$ is defined as the inverse of $b_{ij}$
$$
c \equiv b^{-1}.
$$
Note that the $d_i$ are random variables because they are linear combinations
of other random variables (the $y_k$), but that the $b_{ij}$ (and so
the $c_{ij}$) are combinations of constants, and so are not random variables.
\bigskip
\centerline{\bf Covariances of the Parameters}
\bigskip
We would like to evaluate the covariances of the $a_i$, i.e.,
$\cov(a_i,a_j)$, and derive their errors 
$\sigma(a_i)\equiv\sqrt{\cov(a_i,a_i)}$.  The first step to doing so is
to evaluate the covariances of the $d_i$,
$$
\cov(d_i,d_j) = \cov[y_k \B_{kl} f_i(x_l),y_p\B_{pq} f_j(x_q)] 
= \B_{kl}f_i(x_l)\B_{pq}f_j(x_q)\cov(y_k,y_p).
$$
Then, using the definition of $\C_{kp}$ and successively summing over all
repeated indices,
$$
\cov(d_i,d_j)= \B_{kl}f_i(x_l)\B_{pq}f_j(x_q)\C_{kp} =
\B_{kl}f_i(x_l)\delta_{kq}f_j(x_q)=f_i(x_l)\B_{kl}f_j(x_k)=b_{ij}.
$$
Next, we evaluate the covariance of $a_i$ with $d_j$,
$$
\cov(a_i,d_j) = \cov(c_{im}d_m,d_j)=c_{im}\cov(d_m,d_j)= 
c_{im}b_{mj}=\delta_{ij}.
$$
Finally,
$$
\cov(a_i,a_j) = \cov(c_{im}d_m,a_j)=c_{im}\cov(d_m,a_j)= 
c_{im}\delta_{mj}=c_{ij}.
$$
That is, the $c_{ij}$, which were introduced only to solve for
the $a_i$, now actually turn out to be the covariances of the $a_i$!
In particular,
$$
\sigma(a_i) = \sqrt{c_{ii}}.
$$
This is a very powerful result.  It means that one can figure out the
errors in an experiment, without having any data (just so long as one
knows what data one is planning to get, and what the measurement errors
will be).
\bigskip
\centerline{\bf A simple example}
\bigskip
Let us consider the simple example of a two-parameter straight-line
fit to some data.  In this case,
$$
y_{\rm mod} = a_1f_1(x) + a_2f_2(x) = a_1 + a_2 x,
$$
i.e., $f_1(x)=1$ and $f_2(x) = x$.
Let us also assume that the measurement errors are uncorrelated.
Hence
$$
d_1 = \sum_{k=1}^N {y_k\over \sigma_k^2},\qquad
d_2 = \sum_{k=1}^N {y_k x_k\over \sigma_k^2},
$$
$$
b_{11} = \sum_{k=1}^N {1\over \sigma_k^2},\qquad
b_{12}=b_{21} = \sum_{k=1}^N {x_k\over \sigma_k^2},\qquad
b_{22} = \sum_{k=1}^N {x_k^2\over \sigma_k^2}.
$$
Let us now further specialize to the case for which all the $\sigma_k$
are equal.  Then
$$
b = {N\over \sigma^2}\left(\matrix{1 &\l x \r\cr \l x \r&\l x^2 \r\cr}\right)
$$
where the ``expectation'' signs now mean simply averaging over the
distribution of the independent variable at the times when the observations
actually take place.  This matrix is easily inverted,
$$
c = {\sigma^2\over N(\l x^2 \r - \l x \r^2)}
\left(\matrix{\l x^2 \r & -\l x \r\cr -\l x \r &1}\right).
$$
In particular, the error in the slope is given by
$$
[\sigma(a_2)]^2 = \var(a_2) = c_{22} = {\sigma^2\over N\var(x)}
\rightarrow {12\over N}\,{\sigma^2\over (\Delta x)^2},
$$
where I have used $\var(x)$ as a shorthand for $\l x^2 \r - \l x \r^2$,
and where in the last step I have evaluated this quantity for the special
case of data uniformly distributed over an interval $\Delta x$.
That is, for data with equal errors, the variance of the slope is
equal to the variance of the individual measurements divided by the
variance of the independent-variable distribution, and then divided by $N$.
\bigskip
\centerline{\bf Nonlinear fits}
\bigskip
Consider a more general model,
$$
y_{\rm mod} = F(x;a_1 \ldots a_n),
$$
in which the function $F$ is not linear in the $a_i$, e.g.,
$F(x;a_1,a_2) = \cos(a_1 x)\exp(-a_2 x)$.  The method given above cannot
be used directly to solve for the $a_i$.  Nevertheless, once the minimum
$a_i^0$ is found, by whatever method, one can write
$$
F(x;a_1 \ldots a_n) = F_0 + (a_i - a_i^0)f_i(x) + \ldots
$$
where
$$
f_i(x)\equiv {\p F(x;a_1 \ldots a_n)\over \p a_i}\bigg|_{a_j = a_j^0}.
$$
Then, one can calculate the $b_{ij}$, and so the $c_{ij}$, and thus
obtain the errors. In fact, this formulation can also be used to find the
minimum (using Newton's method), but the description of this approach goes 
beyond the scope of these notes.
\bigskip
\centerline{\bf Expected Value of $\chi^2$}
\bigskip
We evaluate the expected value of $\chi^2$ after it has been
minimized
$$
\l \chi^2\r = \l y_k \B_{kl} y_l \r - 2 \l a_i d_i \r + \l a_i b_{ij} a_j \r
=  \l y_k \B_{kl} y_l \r - \l a_i d_i \r,
$$
where I have used the minimizing condition $d_i = b_{ij} a_j$.
To put this expression in an equivalent form whose physical meaning
is more apparent, it is better to retreat to the original definition,
$$
\l \chi^2 \r = \l [y_k - a_i f_i(x_k)]\B_{kl}[y_l - a_j f_j(x_l)]\r
$$
or
$$
\l \chi^2 \r = \cov([y_k - a_i f_i(x_k)],\B_{kl}[y_l - a_j f_j(x_l)])
+ \l [y_k - a_i f_i(x_k)]\r \B_{kl}\l[y_l - a_j f_j(x_l)]\r.
$$
The first term can be evaluated,
$$
\cov([y_k - a_i f_i(x_k)],\B_{kl}[y_l - a_j f_j(x_l)])
= \B_{kl}\cov(y_k,y_l) - 2\cov(a_i,d_i) + b_{ij}\cov(a_i,a_j) 
$$
$$
= \B_{kl}\C_{kl} - 2\delta_{ii} + b_{ij}c_{ij} = \delta_{kk} - \delta_{ii},
$$
%$$
%\l \chi^2\r = \B_{kl}[\cov(y_k,y_l)+ \l y_k\r\l y_l \r]
%- [\cov(a_i,d_i) + \l a_i \r\l d_i\r] = 
%\B_{kl}\C_{kl}- \delta_{ii} 
%+ \B_{kl}\l y_k\r\l y_l \r - \l a_i \r\l d_i\r,
%$$
using $\cov(a_i,d_j)=\delta_{ij}$.
Since $\delta_{kk}= N$ and $\delta_{ii}=n$
(note that repeated indices still indicate summation), we have
$$
\l \chi^2\r 
%= N -n + \B_{kl}\l y_k\r\l y_l \r - \l a_i \r\l d_i\r
= N-n + \l y_k - a_i f_i(x_k)\r \B_{kl}\l y_l - a_j f_j(x_l)\r.
$$
The last term is composed of the product of the 
expected difference between the model and the data at all pairs of points.
For uncorrelated errors,
$$
\l y_k - a_i f_i(x_k)\r\B_{kl} \l y_l - a_j f_j(x_l)\ \rightarrow
\sum_{k=1}^N {\l y_k - a_i f_i(x_k)\r^2\over \sigma_k^2}.
$$
If the model space spans the physical situation that is being modeled,
then this expected difference is exactly zero.  In this case, 
$\l \chi^2 \r = N-n$, the number of data points less the number of parameters.
If this relation fails, it can only be for one of three reasons:
1) normal statistical fluctuations, 2) misestimation of the errors, 3)
problems in the model.  For Gaussian statistics, one can show
that
$$
\var(\chi^2) = 2(N-n).
$$
So if there are, say 58 data points, and 8 parameters, then $\chi^2$
should be $50\pm 10$.  So if it is 57 or 41, that is ok.  If it is 25
or 108, that is not ok.  If it is 25, the only effect that could have
produced this would be that the errors were overestimated.  If it is 108,
there are two possible causes.  The errors could have been underestimated
or the model could be failing to represent the physical situation.  For
example, if the physical situation causes the data to trace a parabola
(which requires 3 parameters), but the model has only two parameters
(say a 2-parameter straight line fit), then the model cannot adequately
represent the data and the third term will be non-zero and so cause $\chi^2$
to go up.
\bigskip
\centerline{\bf Combining Covariance Matrices}
\bigskip
Suppose one has two sets of measurements, which had been analyzed as
separate $\chi^2$'s, $\chi^2_1$ and $\chi^2_2$.  Minimizing each with 
respect to the $a_i$ yields best-fits $a_i^1$ and $a_i^2$, and associated
vectors and matrices $d_i^1$, $b_{ij}^1$, $c_{ij}^1$, 
$d_i^2$, $b_{ij}^2$, and $c_{ij}^2$.  Now imagine minimizing the sum of
these two $\chi^2$'s with respect to the $a_m$,
$$
0 = {1\over 2}\,{\p (\chi^2_1 + \chi^2_2)\over\p a_m} = 
-(d_m^1 + d_m^2) + (b_{mj}^1 + b_{mj}^2)a_j,
$$
which yields a combined solution,
$$
a_i = c_{ij} d_j,\qquad d_i\equiv d_i^1+d_i^2,\qquad 
b_{ij}\equiv b_{ij}^1+b_{ij}^2, \qquad c\equiv b^{-1}.
$$
But it is not actually necessary to go back to the original $\chi^2$'s.
If you were given the results of the previous analyses, i.e. the best
fit $a_i^1$ and $c_{ij}^1$ and $a_i^2$ and $c_{ij}^2$
for the two separate fits, you could calculate
$b^1 = (c^1)^{-1}$ and $d^1_i = b_{ij}^1 a_j^1$ and similarly for ``2'', and
then directly calculate the combined best fit.
\bigskip
\centerline{\bf Linear Constraints}
\bigskip
Suppose that you have found a best fit to your data, but now
you have obtained some additional information that fixes a linear relation
among your parameters.  Such a constraint can be written
$$
\kappa_i a_i = z.
$$
For example, suppose that your information  was that $a_1$ was actually
equal to 3.  Then $\kappa = (1,0,0, \ldots)$ and $z=3$.  Or suppose
that the information was that $a_1$ and $a_2$ were equal.  Then
$\kappa = (1,-1,0,0,\ldots)$ and $z=0$.  How is the best fit solution
and covariance matrix affected by this constraint?  The answer is
$$
\tilde a_i = a_i^0 - D\alpha_i
\qquad D \equiv {[\kappa_j a_j^0 - z]\over \alpha_p\kappa_p},\qquad
\alpha_i \equiv c_{ij}\kappa_j.
$$
I derive this in a more general context below.  For now just note
that by
$$
\cov(D,D) = {\kappa_i\kappa_j\over(\alpha_p\kappa_p)^2}\cov(a_i^0,a_j^0)
= {\kappa_i\kappa_j\over(\alpha_p\kappa_p)^2}c_{ij} = {1\over\alpha_p\kappa_p},
$$
and
$$
\cov(a_i^0,D) = {\kappa_j\over \alpha_p\kappa_p}\cov(a_i^0,a_j^0)=
{\alpha_i\over \alpha_p\kappa_p},
$$
we obtain,
$$
\tilde c_{ij}\equiv \cov(\tilde a_i,\tilde a_j) = 
\cov(a_i^0,a_j^0) - 2\cov(a_i^0,D)\alpha_j + \cov(D,D)\alpha_i\alpha_j
= c_{ij} - {\alpha_i\alpha_j\over \alpha_p\kappa_p}.
$$
How does imposing this constraint affect the expected value of $\chi^2$?
Substituting $a_i\rightarrow \tilde a_i$, we get all the original terms
plus a $\Delta\l \chi^2 \r$,
$$
\Delta\l \chi^2 \r = 2\alpha_i\l D d_i\r -2\alpha_i b_{ij}\l a_j D\r
+ \alpha_i\alpha_j b_{ij}\l D^2\r =
\alpha_i c_{pj}\kappa_p b_{ij} (\cov(D,D)+\l D \r^2),
$$
or
$$
\Delta\l \chi^2 \r = 1 + {\l D \r^2 \alpha_p\kappa_p}.
$$
That is, if the constraint really reflects reality, i.e. $\l D \r=0$,
then imposing the constraint increases the expected value of $\chi^2$
by exactly unity.  However, if the constraint is untrue, then imposing
it will cause $\chi^2$ to go up by more.
Hence, if the original model space represents reality and the constrained
space continues to do so, then
$$
\l \chi^2 \r = N - n + m,
$$
where $N$ is the number of data points, $n$ is the number of parameters,
and $m$ is the number of constraints.
\bigskip
\centerline{\bf General Constraint and Proof}
\bigskip
As a practical matter, if one has $m>1$ constraints, 
$$
\kappa^k_i a_i = z^k,\qquad (k = 1\ldots m),
$$
these can be
imposed sequentially in a computer program using the above formalism.
However, suppose in a fit of mathematical purity, you decided you 
wanted to impose them all at once.  Then, following Gould \& An 
(2002 ApJ 565 1381), but with slight notation changes, you should
first rewrite,
$$
\chi^2(a_i) = (a_i - a_i^0) b_{ij}(a_j - a_j^0) + \chi^2_0,
$$
where $a_i^0$ is the unconstrained solution and $\chi^2_0$ is the value
of $\chi^2$ for that solution.  At the constrained solution, the
gradient of $\chi^2$ must lie in the $m$ dimensional subspace defined
by the $m$ constraint vectors $\kappa^k_i$.  Otherwise it would be
possible to reduce $\chi^2$ further while still obeying the constraints.
Mathematically,
$$
b_{ij}(a_j - a_j^0) + D^k\kappa^k_i = 0,
$$
where the $D^k$ are $m$ so far undetermined coefficients.  While we do not
yet know these parameters, we can already solve this equation for the
$\tilde a_i$ in terms of them, i.e.,
$$
\tilde a_i = a_i^0 - D^l\alpha^l_i, \qquad \alpha^l_i\equiv c_{ij}\kappa^l_j.
$$
Multiplying this equation by the $\kappa_i^k$ 
yields $m$ equations,
$$
\kappa^k_i\tilde a_i = \kappa_i^k a_i^0 - C^{kl}D^l,\qquad 
C^{kl}\equiv \kappa_i^k \alpha_i^l = \kappa_i^k c_{ij}\kappa_j^l.
$$
Then applying the $m$ constraints gives,
$$
C^{kl} D^l = A^k,\qquad A^k\equiv\kappa^k_ia_i^0 - z^k,
$$
so that the solution for the $D^k$ is
$$
D^k = B^{kl}A^l,\qquad B \equiv C^{-1},
$$
where now the implied summation is over the $m$ constraints
$l=1,\ldots, m$.  Note that,
$$
\cov(A^k,A^l) = \kappa_i^k\kappa_j^l\cov(a_i^0,a_j^0)=
\kappa_i^k\kappa_j^l c_{ij} = C^{kl},
$$
$$
\cov(A^k,D^l) = B^{lq}\cov(A^k,A^q)= B^{lq}C^{kq} = \delta_{kl},
$$
$$
\cov(D^k,D^l) = B^{kq}\cov(A^q,D^l) = B^{kq}\delta_{ql} = B^{kl},
$$
%$$
%\cov(D^k,D^p) = B^{kl}B^{pq}\kappa_i^l\kappa_j^q\cov(a_i^0,a_j^0)
%= B^{kl}B^{pq}C^{lq} = B^{kl}
%$$
and that,
$$
\cov(a_i^0,D^k) = B^{kl}\kappa_j^l\cov(a_i^0,a_j^0) = B^{kl}\alpha^l_i.
$$
Hence,
$$
\cov(\tilde a_i,\tilde a_j) = c_{ij} - 2\alpha_i^k\cov(D^k,a_j^0)
+ \alpha_i^k\alpha_j^l\cov(D^k,D^l)=
c_{ij} - 2\alpha_i^k B^{kl}\alpha^l_j + \alpha_i^k\alpha_j^l B^{kl}.
$$
That is,
$$
\tilde c_{ij} = c_{ij} - \alpha_i^k B^{kl}\alpha_j^l.
$$
Finally, we can evaluate the expected value of $\chi^2$ directly
taking into account the $m$ constraints,
$$\l \chi^2 \r = \B_{kl}C_{kl} - \cov(a_i,d_i) + \cov(A^q,D^q)
+ \B_{kl} \l y_k \r\l y_l \r - \l a_i \r\l d_i \r + \l A^q \r\l D^q\r.
$$
That is,
$$
\l \chi^2 \r = N-n+m + \l y_k - y_{k,\rm mod} \r \B_{kl}
\l y_l - y_{l,\rm mod} \r + \l A^p \r B^{pq}\l A^q \r.
$$
\bigskip
\centerline{\bf Summary}
\bigskip
The $d_i$ (products of the data $y_k$ with the trial functions $f_i(x_l)$)
and the $b_{ij}$ (products of the trial functions with each other), 
$$
d_i \equiv y_k \B_{kl} f_i(x_l),\qquad
b_{ij} \equiv f_i(x_k)\B_{kl} f_j(x_l),\qquad (\B\equiv \C^{-1},\quad
\C_{kl}\equiv \l y_k y_l \r -\l y_k \r \l y_l \r),
$$
are conjugate to the fit parameters $a_i$ and their
associated covariance matrix $c_{ij}$,
$$
\cov(d_i,d_j) = b_{ij},\quad
\cov(a_i,a_j) = c_{ij},\quad
d_i = b_{ij}a_j,\quad
a_i = c_{ij}d_j,\quad
c = b^{-1}.
$$
The parameter errors and covariances $c_{ij}$ can be determined just from
the trial functions, without knowing the data values $y_k$.

Similarly the $A^p$ (products of the unconstrained parameters $a_i^0$
with the constraints $\kappa_i^p$)
and the $C^{pq}$ (products of the constraints with each other), 
$$
A^p \equiv \kappa_i^p a_i^0 - z^p,\qquad
C^{pq} \equiv \kappa_i^p c_{ij}\kappa_j^q,
$$
are conjugate to coefficients of the constrained-parameter 
adjustments $D^p$ and their associated covariance matrix $B^{pq}$,
$$
\cov(A^p,A^q) = C^{pq},\quad
\cov(D^p,D^q) = B^{pq},\quad
A^p = C^{pq}D^q,\quad
D^p = B^{pq}A^q,\quad
B = C^{-1}.
$$
And, while the constrained parameters $\tilde a_i$ of course require 
knowledge of the data, the constrained covariance matrix $\tilde c_{ij}$
does not,
$$
\tilde a_i = a^0_i - D^p\alpha^p_i,\qquad
\tilde c_{ij} = c_{ij} -\alpha^p_i  B^{pq}\alpha^q_j.
$$
Note that the vector adjustments $\alpha_i^p$ have the same relation to 
the constraints $\kappa_i^p$ that the $a_i$ have to the $d_i$,
$$
\alpha_i^p \equiv c_{ij}\kappa_j^p,\qquad
\kappa_i^p = b_{ij}\alpha_j^p
$$
Finally, the expected value of $\chi^2$ is
$$
\l \chi^2 \r = N-n+m + \l y_k - y_{k,\rm mod} \r \B_{kl}
\l y_l - y_{l,\rm mod} \r + \l A^p \r B^{pq}\l A^q \r.
$$
That is, the number of data points, less the number of parameters,
plus the number of constraints, plus two possible additional terms.
The first is zero if the model space spans the system being measured
$(\l y_{k,\rm mod} \r = \l y_k \r)$,
but otherwise is strictly positive.  The second is zero if the constraints
are valid $(\l A^p \r=0)$, but otherwise is strictly positive.
\bigskip
\noindent{\bf acknowledgment:}

I thank David Weinberg and Zheng Zheng for making many helpful suggestions 
including the idea to  place these notes on astro-ph.

\end